\documentclass{article}
\usepackage{spconf,amsmath,graphicx,hyperref}

\usepackage{array}

\usepackage{amsfonts}   
\usepackage{booktabs}  
\usepackage{multirow}  
\usepackage{xcolor}    
\usepackage{enumitem} 

\title{
EEG-SeeGraph: Interpreting Functional Connectivity Disruptions in Dementias via Sparse-Explanatory Dynamic EEG-Graph Learning
}

\name{
Fengcheng Wu\textsuperscript{1}, Zhenxi Song\textsuperscript{1}$^{*}$, Guoyang Xu\textsuperscript{1}, Kaisong Hu\textsuperscript{1}, Zirui Wang\textsuperscript{1}, Yi Guo\textsuperscript{2} and Zhiguo Zhang\textsuperscript{1}
\thanks{
$^{*}$Corresponding author: Zhenxi Song (songzhenxi@hit.edu.cn).
This work was supported by the National Natural Science Foundation of China (Grant No. 62306089) and the Shenzhen Science and Technology Program (Grant Nos. RCBS20231211090800003 and ZDSYS20230626091203008).
}
}

\address{
\textsuperscript{1}Harbin Institute of Technology, Shenzhen, China \\
\textsuperscript{2}Institute of Neurological Diseases, Shenzhen Bay Laboratory, Shenzhen, China
}

\begin{document}
\ninept
\maketitle
\begin{abstract}
% Accurate and interpretable diagnosis of Alzheimer’s Disease (AD) from electroencephalography (EEG) remains challenging, as effective methods must capture evolving temporal dynamics and functional connectivity while maintaining robustness to noisy and non-stationary signals.
% Sentence-1）面对本工作已解决的问题提motivation。
Robust and interpretable dementia diagnosis from noisy, non-stationary electroencephalography (EEG) is clinically essential yet remains challenging.
To this end, we propose \textbf{\textit{SeeGraph}}, a \textit{\textbf{S}parse-\textbf{E}xplanatory dynamic \textbf{E}EG-\textbf{graph}} network that models time-evolving functional connectivity and employs a node-guided sparse edge mask to reveal the connections that drive the diagnostic decision, while remaining robust to noise and cross-site variability.
\textbf{\textit{SeeGraph}} comprises four components: 1) a dual-trajectory temporal encoder that models dynamic EEG with two streams, with node signals capturing regional oscillations and edge signals capturing interregional coupling; 2) a topology-aware positional encoder that derives graph-spectral Laplacian coordinates from the fused connectivity and augments the node embeddings; 3) a node-guided sparse explanatory edge mask that gates the connectivity into a compact subgraph; and 4) a gated graph predictor that operates on the sparsified graph.
The framework is trained with cross-entropy together with a sparsity regularizer on the mask, yielding noise-robust and interpretable diagnoses.
The effectiveness of \textbf{\textit{SeeGraph}} is validated on public and in-house EEG cohorts, including patients with neurodegenerative dementias and healthy controls, under both raw and noise-perturbed conditions.
Its sparse, node-guided explanations highlight disease-relevant connections and align with established clinical findings on functional connectivity alterations, thereby offering transparent cues for neurological evaluation.
\end{abstract}

\section{Introduction}
\label{sec:intro}

% EEG-based diagnostic methods hold significant promise for neurodegenerative disorders such as Alzheimer’s disease (AD), a condition characterized by progressive cognitive decline and the absence of curative therapies. With the accelerating pace of global population aging, China now has the largest number of dementia patients worldwide, accounting for approximately 25\% of all cases and creating a substantial socioeconomic burden~\cite{Jia2020Prevalence}. The strengths of EEG, including low cost, non-invasiveness, and high temporal resolution, make it particularly suitable for AD diagnosis~\cite{Cai2023MBrain}. 
% 1. 介绍本工作相关的研究对象/背景（EEG+dementia）
Electroencephalography (EEG) offers practical utility for screening and monitoring in neurodegenerative dementias such as Alzheimer’s disease (AD) and frontotemporal dementia (FTD)~\cite{Jia2020Prevalence} because it is low-cost, non-invasive, and offers high temporal resolution~\cite{Cai2023MBrain}.
% Nevertheless, its high dimensionality and non-stationary nature pose major challenges for robust deep learning. 
% 2. 介绍本工作相关的目标
Identifying dementia related EEG patterns requires models that can operate on high-dimensional, noisy, and non-stationary signals. It is also valuable that such models generalize across sites\cite{Hata2025EEG} and yield insights that help interpret brain functional disruptions\cite{Jungrungrueang2025Translational}.

% 3. 介绍前人向同一目标努力的工作中，采用的方法手段及其存在的问题（介绍工作时应有一个主线，要么是以sota为中心、要么是以时间发展的角度、要么是以技术类别的角度）但终极目的都是让读者明白本文提出的方法是必要的、是符合发展的、或者是补偿空缺的。
% Although recent advances in dynamic graph networks have enabled the modeling of the brain’s complex topological structure~\cite{4}, existing approaches still suffer from key limitations.
%逻辑是：传统 ML → 普通 DL → 图方法的不足与可解释性问题 → 噪声与跨中心鲁棒性 → 我们的方法动机与一句话概述
Earlier machine learning approaches relied on handcrafted spectral features or shallow classifiers, which are limited in capturing complex spatio-temporal dependencies\cite{Klepl2024GNN}.
Subsequent deep learning models, such as convolutional and recurrent networks, improved representation capacity yet still treated channels as grids or sequences and thus lacked explicit modeling of brain network topology. 
More recent graph-based methods began to encode interregional relationships, but many adopt static or simplified formulations that underutilize temporal evolution\cite{4}\cite{ICASSP2024}.
Since pathological information is frequently embedded in the temporal evolution of network patterns rather than fixed states, existing methods struggle to capture dynamic cross-regional interactions, which undermines generalization and robustness\cite{10,6,8}.

At the same time, many models operate as black boxes, making their decision processes difficult to trace \cite{11,12}. They offer limited insight into the dynamics of functional connectivity and therefore fall short of clinical requirements for transparency and trustworthiness \cite{Yang2021Topology}. Post hoc explanation methods are often unstable or fail to faithfully reflect the model’s reasoning \cite{13,MM2024}.
These limitations highlight the need for inherently interpretable spatio-temporal graph models that provide reliable, clinically meaningful explanations while maintaining strong predictive performance.
Robustness to site variability and acquisition noise also remains insufficient, leaving a gap for approaches that are both reliable and clinically interpretable\cite{ICASPP2025}.

Motivated by dementia pathophysiology that involves disrupted coordinated oscillations and interregional coupling\cite{TBME}, we propose a sparse-explanatory dynamic EEG-graph network, \textit{\textbf{SeeGraph}}, that jointly covers two complementary feature domains: node-level regional oscillatory activity and edge-level amplitude-based interregional coupling. The model forms time-varying brain graphs, learns compact explanatory subnetworks while performing diagnosis, and is designed to improve interpretability and stability under noise and cross-site variability.

% To address these challenges, we propose a dynamic spatio-temporal graph neural network with intrinsic interpretability for AD diagnosis. 
The main contributions of this work are:
\begin{itemize}[leftmargin=*, labelsep=0.5em]
    % \item The model we propose integrates EEG temporal dynamics with node-level functional connectivity, enabling the capture of evolving brain states and pathological transitions.
    \item We introduce \textit{\textbf{SeeGraph}}, an intrinsically explainable dynamic EEG-graph network that models time-varying brain networks for dementia diagnosis while localizing connectivity disruptions.
    % \item We improve the \textbf{robustness of EEG representation learning} through adaptive embeddings and stochastic optimization, enhancing reliability on noisy clinical recordings.
    \item \emph{Quantitatively}, across a public cohort and an in-house clinical cohort, \textit{\textbf{SeeGraph}} achieves \emph{state-of-the-art} accuracy under both raw and noise-perturbed EEG, demonstrating cross-site robustness.
    % \item We propose an \textbf{interpretable connectivity mechanism} inspired by the Information Bottleneck, where a stochastic attention mask highlights compact and discriminative patterns, yielding clinically meaningful explanations alongside accurate AD diagnosis.
    \item \emph{Qualitatively}, \textit{\textbf{SeeGraph}} yields sparse, node-guided edge-level explanations that identify disease-relevant functional connectivity consistent with clinical evidence.
\end{itemize}

\begin{figure*}[t]
    \centering
    \centering
    \includegraphics[width=1.0\textwidth]{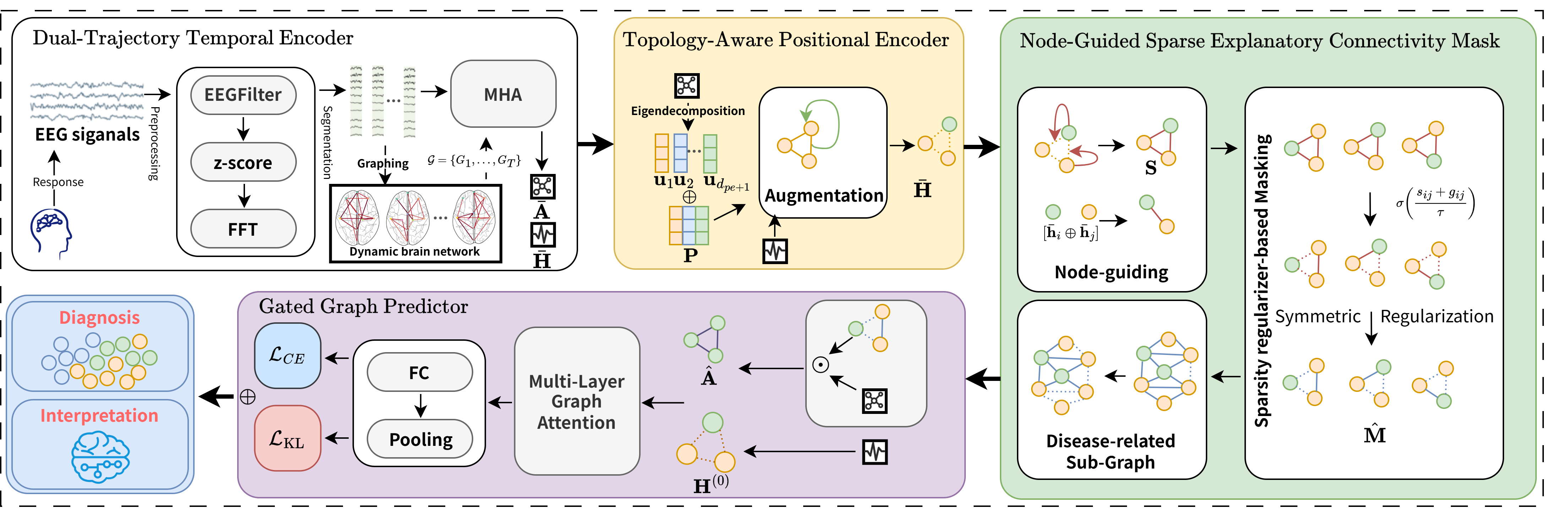} 
    % \caption{Overall Architecture of  \textbf{\textit{SeeGraph}}}
    \caption{Architecture of \textbf{\textit{SeeGraph}}: dynamic EEG graphs are constructed, node and connectivity trajectories are temporally encoded, Laplacian positional encodings augment node embeddings, and a node-guided sparse mask yields an explanatory subgraph for a gated graph predictor. The objective combines diagnosis cross-entropy with a sparsity regularizer on the edge mask to learn compact, interpretable connectivity.}
    \label{model1}
\end{figure*}

\section{METHOD}
\label{sec:method}
% We propose  \textbf{\textit{SeeGraph}}, a spatio-temporal graph neural network with built-in interpretability for dynamic brain network analysis. 
The overall architecture of \textbf{\textit{SeeGraph}}, an intrinsically explainable dynamic EEG-graph network, is illustrated in Fig.~\ref{model1}.
%  \textbf{\textit{SeeGraph}} first transforms multi-channel EEG signals into a sequence of temporal dynamic graphs. It then adopts a dual-path architecture that jointly learns discriminative representations for classification and identifies an interpretable network consisting of key connections, thereby achieving a balance between predictive performance and interpretability.
% ###### 重点看以下内容，这是本文方法模块阐述，方法特色是实现本文贡献的途径（全文表述围绕）#####
First, the \textbf{dual-trajectory temporal encoder} introduces two feature domains that together cover regional oscillatory activity and interregional coupling. Node trajectories carry frequency domain spectra obtained by FFT, while edge trajectories carry Pearson correlation computed from amplitude time series; the two streams share parameters in the dual-trajectory temporal encoder to improve data efficiency and robustness under noisy, limited clinical EEG.
Second, the \textbf{topology-aware positional encoder} derives graph-spectral Laplacian coordinates from the fused edge connectivity and appends them to the node embeddings, injecting task-relevant topological context into the node domain.
Third, the \textbf{node-guided sparse explanatory edge mask} conditions sparsification on pairwise node embeddings instead of raw edge weights, which avoids circular dependence on the connectivity being gated, prevents mask degeneration, and yields faithful, compact explanatory subgraphs.
Fourth, the \textbf{gated graph predictor} runs graph attention on the sparsified connectivity and aggregates node representations into a graph-level embedding for diagnosis. The training objective couples cross-entropy that pulls the learned edge mask toward a low retention Bernoulli prior, implementing an information bottleneck style constraint.
% ###方法图应与上述文字有关键词对应，可标公式对应的符号，让人一目了然方法是如何运作的###

\subsection{Dual-Trajectory Temporal Encoder}
\label{ssec:graph_construction}

% As illustrated in Fig.~\ref{model1}, prior to being fed into the model, each EEG channel is Z-score normalized by subtracting its mean and dividing by its standard deviation. We represent a multi-channel EEG sequence with $N$ channels and $T$ time windows as a graph sequence $\mathcal{G} = \{ G_1, G_2, \dots, G_T \}$, where each $G_t = (V, E_t, \mathbf{X}_t, \mathbf{A}_t)$ is defined as:
As illustrated in Fig.~\ref{model1}, we z-score each channel and segment the $N$-channel EEG into $T$ windows to form a dynamic graph sequence $\mathcal{G}=\{G_t\}_{t=1}^{T}$. Each graph $G_t=(V,E_t,\mathbf{X}_t,\mathbf{A}_t)$ is defined so that $V$ indexes the $N$ EEG channels (node $v_i$ corresponds to one channel), $E_t$ collects the functional couplings in window $t$, $\mathbf{X}_t\in\mathbb{R}^{N\times d}$ stacks node features with row $i$ given by $\mathbf{x}_{i,t}\in\mathbb{R}^{d}$ encoding the activity of channel $i$ at time $t$, and $\mathbf{A}_t\in\mathbb{R}^{N\times N}$ is the adjacency whose entry $a_{ij,t}$ quantifies the functional strength between $v_i$ and $v_j$ at time $t$.
% \begin{itemize}[leftmargin=*, labelsep=0.5em]
%     \item \textbf{Nodes ($V$):} Each node $v_i \in V$ corresponds to an EEG channel, with $|V|=N$.
%     \item \textbf{Connectivity ($E_t$):} Functional couplings among channels at time window $t$.
%     \item \textbf{Node Features ($\mathbf{X}_t$):} $\mathbf{X}_t \in \mathbb{R}^{N \times d}$, where $i$-th row $\mathbf{x}_{i,t}\in\mathbb{R}^{d}$ encodes channel $i$ activity in window $t$.
%     \item \textbf{Adjacency ($\mathbf{A}_t$):} $\mathbf{A}_t \in \mathbb{R}^{N \times N}$, where $a_{ij,t}$ quantifies the functional strength between $v_i$ and $v_j$ at time $t$.
% \end{itemize}
Unlike static methods with a fixed adjacency, we construct $\mathbf{A}_t$ for each time window to preserve brain network dynamics.

% \noindent\textbf{Feature Encoding.}
The dual-stream temporal encoder introduces two complementary feature domains that jointly capture regional oscillatory activity and inter-regional coupling. Specifically, the continuous EEG signal is segmented into short, overlapping time windows. For each window $t$, we extract two types of features:

\begin{itemize}[leftmargin=*, labelsep=0.5em]
    \item \textbf{Node Trajectory:} 
    % For each channel, we apply the Fast Fourier Transform (FFT) to obtain its spectral representation. The resulting frequency-domain features are used as the node attributes $\mathbf{X}_t \in \mathbb{R}^{N \times d}$, where $\mathbf{x}_{i,t} \in \mathbb{R}^{d}$ encodes the oscillatory activity of channel $i$ at time $t$.
    or each channel $i$ and time window $t$, we apply the Fast Fourier Transform (FFT) to the windowed signal to obtain a spectral vector $\mathbf{x}_{i,t}\in\mathbb{R}^{d}$. Stacking these vectors over time yields the node trajectory $\mathbf{X}_i=[\mathbf{x}_{i,1};\dots;\mathbf{x}_{i,T}]\in\mathbb{R}^{T\times d}$, which summarizes the oscillatory activity of channel $i$ across windows. Stacking $\mathbf{X}_i$ across channels produces the tensor $\mathbf{X}_{1:T}\in\mathbb{R}^{T\times N\times d}$ that collects all node tokens for the $T$ windows.
    \item \textbf{Edge Trajectory:} 
    % Based on the amplitude time series of EEG channels, we compute the absolute Pearson correlation between each pair of channels, thereby constructing a weighted symmetric adjacency matrix $\mathbf{A}_t \in \mathbb{R}^{N \times N}$, where $a_{ij,t}$ denotes the functional coupling strength between channel $i$ and channel $j$ at time $t$.
    Within each window $t$, we compute the absolute Pearson correlation between the amplitude time series of each pair of channels $(i,j)$, producing a weighted, symmetric adjacency matrix $\mathbf{A}_t\in\mathbb{R}^{N\times N}$ with entries $a_{ij,t}$ and zero diagonal. Stacking $\{\mathbf{A}_t\}_{t=1}^{T}$ over time yields the edge trajectory tensor $\mathbf{A}_{1:T}\in\mathbb{R}^{T\times N\times N}$. For a specific pair $(i,j)$, the temporal trajectory $\mathbf{a}_{ij,1:T}=[a_{ij,1},\dots,a_{ij,T}]^\top\in\mathbb{R}^{T}$ represents the evolution of interregional coupling across windows. We adopt absolute Pearson correlation for its parameter-free formulation and robustness under short-window and noisy EEG conditions.
    % \textcolor{red}{
    % We adopt absolute Pearson correlation for its parameter-free formulation and robustness under short-window and noisy EEG conditions.
    % }

\end{itemize}

After processing all $T$ time windows, we construct a sequence of dynamic graphs $\mathcal{G}=\{G_t\}_{t=1}^{T}$, whose topology evolves.
We then apply multi-head self-attention (MHA)\cite{Attention} along the temporal axis independently to each node trajectory and each connectivity trajectory, capturing long-range temporal dependencies.
% and highlighting key functional connectivity patterns, thereby facilitating subsequent classification and interpretation.
Given an input sequence $X=\{x_1,\dots,x_T\}$, MHA computes weighted combinations of projected queries $\mathbf{Q}_k$, keys $\mathbf{K}_k$, and values $\mathbf{V}_k$, followed by an output projection $\mathbf{W}^O$~\cite{Attention}.
Specifically, two feature streams share parameters within MHA s, improving data efficiency and robustness under noisy and limited clinical EEG conditions. 
% For an input sequence \(X=\{x_1,\dots,x_T\}\) with \(x_t\in\mathbb{R}^d\),
\begin{align}
\mathrm{MHA}(X) &= \mathrm{Concat}(\mathrm{head}_1,\dots,\mathrm{head}_H)\,\mathbf{W}^O,\\
\mathrm{head}_k &= \mathrm{softmax}\!\left(\frac{\mathbf{Q}_k\mathbf{K}_k^\top}{\sqrt{d_k}}\right)\mathbf{V}_k,
\end{align}
% where \(\mathbf{Q}_k,\mathbf{K}_k,\mathbf{V}_k\) are linear projections of \(X\) and \(d_k\) is the key dimension. This produces time-aware embeddings that capture long-range dependencies and highlight evolving functional connectivity, benefiting both classification and interpretation.

% In this dual-trajectory temporal encoder, temporal dependencies are fused into compact representations. The resulting node embeddings $\bar{\mathbf{H}} \in \mathbb{R}^{N \times D}$ encode the temporally integrated activity of each channel, while the connectivity embeddings $\bar{\mathbf{A}} \in \mathbb{R}^{N \times N}$ summarize the evolving functional connectivity across time. The node embeddings are averaged across time to obtain a time-integrated representation, capturing the overall activity of each node. For the graph connectivity, we select the connectivity embeddings from the final timestep, which captures the most recent interactions between nodes.

The dual-trajectory temporal encoder fuses temporal dependencies into compact representations. It yields node embeddings $\bar{\mathbf H}\in\mathbb{R}^{N\times D}$ that summarize each channel’s activity over time and a fused connectivity matrix $\bar{\mathbf A}\in\mathbb{R}^{N\times N}$ that captures the evolution of functional coupling. 
We compute $\bar{\mathbf H}$ by averaging the node time tokens, and we form $\bar{\mathbf A}$ by taking the last time step of the edge stream, which emphasizes the most recent interregional interactions.
We compute $\bar{\mathbf H}$ by averaging the node time tokens, and we form $\bar{\mathbf A}$ by taking the final attention-aggregated token of the edge stream, which emphasizes the most recent interregional interactions.

\subsection{Topology-Aware Positional Encoder}
\label{ssec:st_framework}
% We further incorporate structural information through a \textbf{Topology-Aware Positional Encoder}. 
This encoder derives Laplacian eigenvector coordinates from the fused adjacency and concatenates them with the node embeddings, thereby injecting topology-relevant positional cues into the node domain\cite{15}. Given the temporally fused connectivity \(\bar{\mathbf{A}}\in\mathbb{R}^{N\times N}\), let \(\mathbf{D}=\mathrm{diag}(\bar{\mathbf{A}}\mathbf{1})\) be its degree matrix and define the (symmetric) normalized Laplacian.
\begin{equation}
\mathbf{L}=\mathbf{I}-\mathbf{D}^{-\frac{1}{2}}\bar{\mathbf{A}}\mathbf{D}^{-\frac{1}{2}}.
\end{equation}
% The eigendecomposition \(\mathbf{L}=\mathbf{U}\boldsymbol{\Lambda}\mathbf{U}^\top\) yields Laplacian positional encodings \(\mathbf{P}=[\mathbf{p}_1^\top,\dots,\mathbf{p}_N^\top]^\top\in\mathbb{R}^{N\times d_{\text{pe}}}\) . 
Taking the eigendecomposition of $\mathbf L$ yields $\mathbf L=\mathbf U\,\boldsymbol{\Lambda}\,\mathbf U^\top$,
where $\mathbf U = [\,\mathbf u_1,\ldots,\mathbf u_N\,]$ is the eigenvector matrix with columns 
$\mathbf u_i \in \mathbb{R}^N$, and $\boldsymbol{\Lambda}=\mathrm{diag}(\lambda_1,\ldots,\lambda_N)$ are the eigenvalues in ascending order.
The Laplacian positional encodings $\mathbf P$ are then constructed by stacking the $d_{\text{pe}}$ eigenvectors
from $\mathbf U$ that correspond to the smallest nonzero eigenvalues:
\begin{equation}
\mathbf P \;=\; [\,\mathbf u_2,\ldots,\mathbf u_{d_{\text{pe}}+1}\,] \in \mathbb{R}^{N\times d_{\text{pe}}}.
\end{equation}
We then augment graph connectivity embeddings via concatenation:
% We then form the \emph{augmented} node embeddings by concatenation:
\begin{equation}
\bar{\mathbf H} \leftarrow [\,\bar{\mathbf H}\,\|\,\mathbf P\,] \in \mathbb{R}^{N\times (D+d_{\text{pe}})}. 
\end{equation}
By embedding both local activity and global structure, the model gains a more holistic understanding of brain dynamics.

% Finally, the temporally fused connectivity $\bar{\mathbf A}$ derived from dynamic graphs is gated by the explanatory mask to obtain the explanatory network used for message passing:
% \begin{equation}
% \hat{\mathbf A} = \tilde{\mathbf M} \odot \bar{\mathbf A}. \tag{5}
% \end{equation}

\subsection{Node-Guided Sparse Explanatory Connectivity Mask}
\label{sssec:interpretable_learning}

% We adopt a dual-path design inspired by the Information Bottleneck: an \textbf{Extractor} learns a sparse \emph{explanatory structural brain network}, and a \textbf{Predictor} performs classification conditioned on the brain network.

% \noindent\textbf{Extractor (Informative connectivity patterns Identification).}
% The Extractor, by $\phi$, 
This module is designed to identify a sparse and informative network by learning a stochastic connectivity mask\cite{StochasticAttention}. It operates in an \emph{connectivity-wise} manner, where the importance of each potential connectivity is computed directly from the embeddings of its endpoint nodes.

Specifically, for each potential connectivity $(v_i, v_j)$, the Extractor takes the corresponding time-aggregated node embeddings $\bar{\mathbf{h}}_i$ and $\bar{\mathbf{h}}_j$ from the previous stage. These two embeddings are concatenated to form a feature vector $[\bar{\mathbf{h}}_i \oplus \bar{\mathbf{h}}_j]$ that uniquely represents the connectivity. This vector is then passed through a shared Multi-Layer Perceptron (MLP) to produce a single logit $s_{ij}$ quantifying the connectivity's relevance. Repeating this for all pairs yields the full \emph{connectivity-logit matrix} $\mathbf{S} \in \mathbb{R}^{N \times N}$.

From these logits, a differentiable \emph{binary-concrete} (Gumbel-Sigmoid) mask $\mathbf{M} \in (0, 1)^{N \times N}$ is sampled:
\begin{equation}
m_{ij}=\sigma\!\left(\frac{s_{ij}+g_{ij}}{\tau}\right), \quad g_{ij}\sim \mathrm{Logistic}(0,1),
\end{equation}
where $\tau > 0$ is a temperature parameter that is annealed during training. To ensure the resulting network is undirected, the mask is made symmetric and its diagonal is zeroed out:
\begin{equation}
\tilde{\mathbf{M}}=\mathrm{SymmZeroDiag}(\mathbf{M})=\frac{\mathbf{M}+\mathbf{M}^\top}{2}-\mathrm{diag}\!\big(\mathrm{diag}(\mathbf{M})\big).
\end{equation}
To encourage sparsity and prevent the model from selecting all connectivity, a KL divergence penalty aligns the mask distribution with a Bernoulli prior\cite{Softmax,B2} of mean $r$:
\begin{equation}
\mathcal{L}_{\mathrm{KL}}=\frac{1}{N(N-1)}\sum_{\substack{i\neq j}}\!\!\Big[m_{ij}\log\frac{m_{ij}}{r+\epsilon}+(1-m_{ij})\log\frac{1-m_{ij}}{1-r+\epsilon}\Big],
\end{equation}
where $\epsilon$ ensures numerical stability. Sparsification is performed based on pairwise node embeddings rather than raw edge weights, which avoids circular dependence on the connectivity being gated, prevents mask degeneration, and yields reliable and compact explanatory brain network.

\subsection{Gated Graph Predictor}
To form the input graph structure for classification, we construct a gated adjacency matrix by applying the learned explanatory mask to the temporally fused connectivity:
\begin{equation}
\hat{\mathbf{A}} = \tilde{\mathbf{M}} \odot \bar{\mathbf{A}},
\end{equation}
where \(\bar{\mathbf{A}}\in\mathbb{R}^{N\times N}\) is the dynamic connectivity matrix aggregated over time. This gating mechanism selectively retains the most relevant connections, and The resulting \(\hat{\mathbf{A}}\) is used in subsequent graph learning steps to guide the learning process in the network for diagnosis.

% \noindent\textbf{Predictor (Classification on brain network).}
Let \(\mathbf{H}^{(0)}=[\,\bar{\mathbf H}\,\|\,\mathbf P\,]\in\mathbb{R}^{N\times(D+d_{\text{pe}})}\) be the initial node-feature matrix.
A multi-layer GAT\cite{GAT} runs on \(\hat{\mathbf A}\), updating rows of \(\mathbf{H}^{(l)}\) as
\begin{equation}
\mathbf{H}^{(l+1)}_{i:}
=\sigma\!\left(\sum_{j\in\mathcal{N}_{\hat{\mathbf A}}(i)}
\hat{a}_{ij}\,\alpha^{(l)}_{ij}\,\mathbf{W}^{(l)}\,\mathbf{H}^{(l)}_{j:}\right),
\end{equation}
where \(\hat{a}_{ij}\) is the \((i,j)\)-entry of the explanatory adjacency \(\hat{\mathbf A}\), 
\(\alpha^{(l)}_{ij}\) are GAT attention coefficients, and \(\mathbf{W}^{(l)}\) are learnable weights.After graph-level pooling, we obtain a graph representation for classification and train the model by minimizing the objective:
\begin{equation}
\mathcal{L}=\mathcal{L}_{\mathrm{CE}} + \lambda_{\mathrm{KL}}\mathcal{L}_{\mathrm{KL}},
\end{equation}
The training objective combines cross-entropy loss with a term that pulls the learned edge mask toward a low retention Bernoulli prior, thereby implementing an information bottleneck-style constraint. Here, $\lambda_{\mathrm{KL}}$ is a hyperparameter that balances the trade-off between predictive accuracy and the sparsity of the explanation.

This mechanism distills the dynamic brain network into a sparse, clinically meaningful subgraph that makes explicit which connections drive the diagnosis, thereby enhancing transparency and robustness by filtering spurious relations.

\begin{table*}[t]
  \centering
  \footnotesize
  \caption{Performance comparison on the SZPH and AHEPA datasets under normal and noisy conditions}
  \label{tab:my_new_results}
  % \begin{tabular}{l|ccc|ccc|ccc|ccc}
  \begin{tabular}{>{\centering\arraybackslash}p{1.5cm}|ccc|ccc|ccc|ccc}

    \hline
    \multirow{3}{*}{\textbf{Method}} & \multicolumn{6}{c|}{\textbf{SZPH Dataset (\textit{China})}} & \multicolumn{6}{c}{\textbf{AHEPA Dataset (\textit{Greece})}} \\
    \cline{2-7} \cline{8-13} 
    & \multicolumn{3}{c|}{Raw (no added noise)} & \multicolumn{3}{c|}{Noisy} & \multicolumn{3}{c|}{Raw (no added noise)} & \multicolumn{3}{c}{Noisy} \\
    \cline{2-4} \cline{5-7} \cline{8-10} \cline{11-13}
    & ACC & AUROC & F1 & ACC & AUROC & F1 & ACC & AUROC & F1 & ACC & AUROC & F1 \\
    \hline
    BIOT\cite{Biot} & 0.765 & 0.820 & 0.761 & 0.621 & 0.767 &  0.607 & 0.812 & 0.937 & 0.798 & 0.733 & 0.786 & 0.712 \\
    DCRNN\cite{Tang2021SSGNN} & 0.760 & 0.819 & 0.779 & 0.661 & 0.740 & 0.737 & 0.738 & 0.819 & 0.719 & 0.651 & 0.734 & 0.657 \\
    EGCN\cite{Pareja2020EvolveGCN} & 0.734 & 0.726 & 0.718 & 0.645 & 0.643 & 0.660 & 0.648 & 0.769 & 0.614 & 0.602 & 0.736 & 0.522 \\
    CNN-LSTM & 0.770 & 0.804 & 0.772 & 0.723 & 0.731 & 0.695 & 0.635 & 0.752 & 0.556 & 0.592 & 0.631 & 0.542 \\
    STGCN\cite{Shao2022STGNN} & 0.772 & 0.831 & 0.761 & 0.743 & 0.785 & 0.717 & 0.645 & 0.778 & 0.628 & 0.623 & 0.758 & 0.603 \\
    MATT\cite{Pan2022MAtt} & 0.613 & 0.653 & 0.630 & 0.583 & 0.626 & 0.571 & 0.669 & 0.729 & 0.627 & 0.635 & 0.701 & 0.589 \\
    \hline
    \textbf{\textbf{\textit{SeeGraph}}} & \textbf{0.874} & \textbf{0.951} & \textbf{0.875} & \textbf{0.848} & \textbf{0.926} & \textbf{0.849} & \textbf{0.841} & \textbf{0.953} & \textbf{0.834} & \textbf{0.827} & \textbf{0.937} & \textbf{0.822} \\
    \hline
  \end{tabular}
\end{table*}

\vspace{-0.2cm}
\begin{table*}[t]
  \centering
  \footnotesize
  \caption{Diagnostic utility of EEG frequency bands for dementia diagnosis across two cohorts}
  \label{tab:freq_band_comparison}
  \begin{tabular}{>{\centering\arraybackslash}p{1.5cm}|ccc|ccc|ccc|ccc}
    \hline
    \multirow{3}{*}{\textbf{Bands}} 
      & \multicolumn{6}{c|}{\textbf{SZPH Dataset (\textit{China})}} 
      & \multicolumn{6}{c}{\textbf{AHEPA dataset (\textit{Greece})}} \\
    \cline{2-7} \cline{8-13}
      & \multicolumn{3}{c|}{Raw (no added noise)} & \multicolumn{3}{c|}{Noisy}
      & \multicolumn{3}{c|}{Raw (no added noise)} & \multicolumn{3}{c}{Noisy} \\
    \cline{2-4} \cline{5-7} \cline{8-10} \cline{11-13}
      & ACC & AUROC & F1 & ACC & AUROC & F1 & ACC & AUROC & F1 & ACC & AUROC & F1 \\
    \hline
    $\delta$ (Delta) & 0.693 & 0.795 & 0.716 & 0.663 & 0.765 & 0.664 & NAN & NAN & NAN & NAN & NAN & NAN \\
    $\theta$ (Theta) & 0.702 & 0.867 & 0.621 & \textbf{0.815} & \textbf{0.830} & \textbf{0.804} & 0.739 & 0.897 & 0.719 & 0.725 & 0.876 & 0.710 \\
    $\alpha$ (Alpha) & 0.792 & 0.817 & 0.773 & 0.729 & 0.816 & 0.745 & \textbf{0.881} & \textbf{0.969} & \textbf{0.876} & \textbf{0.832} & \textbf{0.937} & \textbf{0.851} \\
    $\beta$ (Beta)   & \textbf{0.828} & \textbf{0.887} & \textbf{0.818} & 0.785 & 0.862 & 0.764 & 0.827 & 0.948 & 0.820 & 0.789 & 0.895 & 0.763 \\
    $\gamma$ (Gamma) & 0.708 & 0.819 & 0.748 & 0.692 & 0.770 & 0.660 & 0.681 & 0.859 & 0.676 & 0.676 & 0.815 & 0.663 \\
    \hline
  \end{tabular}
\end{table*}

% \begin{table*}[t]
%   \centering
%   \caption{Impact of EEG frequency bands on AD diagnosis performance across two datasets}
%   \label{tab:freq_band_comparison}
%   \begin{tabular}{l | ccc | ccc}
%     \toprule
%     \multirow{2}{*}{\textbf{Frequency Band}} 
%     & \multicolumn{3}{c|}{\textbf{SZPH Dataset}} 
%     & \multicolumn{3}{c}{\textbf{AHEPA dataset}} \\
%     \cmidrule(lr){2-4} \cmidrule(lr){5-7}
%     & ACC & AUROC & F1 & ACC & AUROC & F1 \\
%     \midrule
%     $\delta$ (Delta) & 0.6931 & 0.7949 & 0.7156 & NAN & NAN & NAN \\
%     $\theta$ (Theta) & 0.7017 & 0.8669 & 0.6212 & 0.7387 & 0.8969 & 0.7193 \\
%     $\alpha$ (Alpha) & 0.7915 & 0.8170 & 0.7729 & \textbf{0.8808} & \textbf{0.9692} & \textbf{0.8762} \\
%     $\beta$ (Beta)   & \textbf{0.8279} & \textbf{0.8865} & \textbf{0.8177} & 0.8273 & 0.9476 & 0.8203 \\
%     $\gamma$ (Gamma) & 0.7075 & 0.8194 & 0.7480 & 0.6806 & 0.8589 & 0.6755 \\
%     \bottomrule
%   \end{tabular}
% \end{table*}

\section{EXPERIMENTS AND RESULTS}
\label{sec:experiments}

\subsection{Experimental Setup}
\label{ssec:EXPERIMENTAL SETUP}
\noindent\textbf{Datasets.} 
% We employed two EEG datasets. The first, \emph{AHEPA dataset}, is a publicly available resting-state EEG collection \cite{EEGDataset}, hereafter referred to as the AHEPA dataset (the dataset is named after the hospital due to its collection and contribution to the research). It consists of 88 subjects, including 36 individuals diagnosed with Alzheimer’s disease(AD), 23 individuals diagnosed with frontotemporal dementia, and 29 healthy control subjects. Recordings were acquired with eyes closed using a 19-channel system (10–20 standard, 500 Hz). We considered a three-class classification task, and evaluation metrics were computed in a macro-averaged manner across the three classes. The second dataset was collected at Shenzhen People’s Hospital (SZPH), comprising clinical EEG data from AD patients and healthy controls (HC). After preprocessing (filtering, artifact removal, segmentation) and diagnostic screening, 50 subjects remained (30 AD, 20 HC). Signals were acquired with a 64-channel 10–20 system.
We used two EEG datasets. (i) AHEPA: a public resting-state EEG cohort collected at AHEPA University Hospital \cite{EEGDataset}. It includes 88 participants: 36 with AD, 23 with FTD, and 29 healthy controls (HC). Recordings were acquired in the eyes-closed condition with a 19-channel 10–20 system at 500 Hz. We treat this as a three-class classification task and report macro-averaged metrics across classes. (ii) SZPH: a clinical cohort from Shenzhen People’s Hospital comprising EEG from patients with AD and HC. A total of 50 participants remained (30 AD, 20 HC) after diagnostic screening. Recordings used a 64-channel 10–20 system.

% \noindent\textbf{Metrics.} We evaluated model performance using Accuracy, AUROC, and F1 score. 

% Accuracy measures the proportion of correctly identified patients and healthy controls. AUROC assesses the model’s discriminative ability across thresholds, while F1 emphasizes the optimal balance between precision and recall for the classification task.

% \noindent\textbf{Model training.} Experiments were conducted using PyTorch on an NVIDIA RTX 3090 GPU. The SZPH dataset, newly collected for this study, is incorporated alongside the publicly available AHEPA dataset to validate the effectiveness of the proposed approach. For sample construction, we used an 8-s window, with segment lengths of 250 points for SZPH and 500 points for AHEPA dataset.

\noindent\textbf{Implementation details.}
We implemented \textit{\textbf{SeeGraph}} in PyTorch and ran all experiments on an NVIDIA RTX 3090 GPU.
Evaluation was conducted on the in-house SZPH cohort and the public AHEPA cohort.
% EEG was segmented into fixed 8-s windows, with 250 and 500 time steps per window for SZPH and AHEPA, respectively.
% \textcolor{red}{
% All experiments are conducted in a dataset-specific manner without cross-dataset training or testing: each dataset is processed under its native montage, treating channels as graph nodes without channel alignment. Differences in sampling rates and channel counts are handled by consistent band-limiting, spectral feature extraction, and z-score normalization computed on training subjects only.
% }
Experiments were conducted independently per dataset. Differences in sampling rates and channel counts were handled via consistent band-limiting and spectral feature extraction.
% }

\subsection{Experimental Results}

To ensure 
% \textcolor{red}{
fair comparison in our subject-independent clinical setting, we focus on representative dynamic GNN and temporal baselines rather than externally pretrained EEG foundation models. Baselines include: dynamic GNNs such as EvolveGCN-O~\cite{Pareja2020EvolveGCN}, DCRNN ~\cite{Tang2021SSGNN}, and STGCN~\cite{Shao2022STGNN}, and temporal models including BIOT~\cite{Biot}, CNN-LSTM, and MAtt~\cite{Pan2022MAtt}.
% }
% \textcolor{red}{
All experiments use a subject-independent 80/20 train–test split at the subject level.
% }

% We compared \textbf{\textit{SeeGraph}} with representative baselines, including dynamic GNNs such as EvolveGCN-O~\cite{Pareja2020EvolveGCN}, DCRNN ~\cite{Tang2021SSGNN}, and STGCN~\cite{Shao2022STGNN}, and temporal models including BIOT~\cite{Biot}, CNN-LSTM, and MAtt~\cite{Pan2022MAtt} (manifold attention over SPD, Euclidean, and Stiefel spaces).

% Each dataset was split into stratified training and test sets in an 80/20 ratio. 

As shown in Table~\ref{tab:my_new_results}, \textbf{\textit{SeeGraph}} achieves the best metrics on both the SZPH and AHEPA datasets, 
confirming the benefits of the two-stream temporal encoding and node-guided sparse edge gating. 
% validating the effectiveness of the stochastic attention mechanism and frequency priors in AD decoding. 
The same table reports robustness under additive Gaussian noise (zero mean, standard deviation 0.3) injected into the input signals.
Under these noisy conditions, \textbf{\textit{SeeGraph}} exhibits only minor performance degradation, demonstrating robustness, effective modeling of long-range temporal dependencies, and generalization across varying data quality.

\vspace{-0.8cm}
\begin{table}[h]
  \centering
  \footnotesize
  \caption{Ablation study of \textbf{\textit{SeeGraph}}}
  % \vspace{0.5em}  
  \label{tab:ablation_normal}
  \begin{tabular}{l|ccc|ccc}
    \hline
    \multirow{2}{*}{\textbf{Method}} & \multicolumn{3}{c|}{\textbf{SZPH}} & \multicolumn{3}{c}{\textbf{AHEPA}} \\
    \cline{2-4} \cline{5-7}
     & ACC & AUROC & F1 & ACC & AUROC & F1 \\
    \hline
    \textbf{\textbf{\textit{SeeGraph}}} & \textbf{0.874} & \textbf{0.951} & \textbf{0.875} & \textbf{0.841} & \textbf{0.953} & \textbf{0.834} \\
    \hline
    {w/o C-wise}  & 0.855 & 0.938 & 0.862 & 0.823 & 0.904 & 0.805 \\
    {w/o PE}  & 0.837 & 0.905 & 0.841 & 0.824 & 0.918 & 0.829 \\
    {w/o SR}  & 0.770 & 0.864 & 0.761 & 0.812 & 0.937 & 0.798 \\
    {w/o FFT} & 0.758 & 0.842 & 0.772 & 0.784 & 0.924 & 0.768 \\
    \hline
  \end{tabular}
\end{table}

For scientific validation and clinical relevance, we conduct a band-wise analysis to quantify each EEG band’s contribution to dementia diagnosis. Results (Table \ref{tab:freq_band_comparison}) shows that $\alpha$, $\beta$, and $\delta$ carry the strongest diagnostic signal, $\theta$ is moderate, and $\gamma$ is comparatively weak. In the AHEPA cohort, the $\delta$-band is unavailable after their original preprocessing, yielding NA entries. These findings are consistent with clinical evidence of $\alpha$ attenuation, $\beta$ suppression, and $\delta$ elevation in dementia, thereby reinforcing the plausibility and interpretability of the learned features.

% We conducted ablation experiments to assess the contribution of each module in  \textbf{\textit{SeeGraph}}. The results in Table~\ref{tab:ablation_normal} show a significant performance drop after removing any component, highlighting their complementary roles in enhancing representation learning. Specifically, removing connectivity-wise modeling (w/o C-wise) , which provides the direct functional connectivity weights to the predictor, leads to a decrease in multiple metrics, emphasizing the importance of capturing interrelations between EEG channels. Similarly, removing frequency-domain feature extraction (w/o FFT) causes a notable decline in performance, underlining the value of frequency-domain information in understanding brain dynamics. Eliminating the sparsity regularizer (w/o SR) leads to a substantial drop in AUROC, demonstrating its critical role in learning relevant features. Finally, removing positional encoding (w/o PE) slightly decreases F1 score, stressing the importance of topological encoding for classification accuracy and interpretability. These results confirm the indispensability of each module in  \textbf{\textit{SeeGraph}} for improving both performance and interpretability.

We conducted ablation experiments to assess the contribution of each module in \textbf{\textit{SeeGraph}}. As shown in Table~\ref{tab:ablation_normal}, removing any single component degrades performance, indicating complementary contributions to representation learning. Removing the connectivity stream (w/o C-wise) lowers all metrics, underscoring the importance of modeling inter-channel relationships.  Discarding positional encodings (w/o PE) also reduces F1, showing that topological cues aid both accuracy and interpretability. Turning off the sparsity regularizer on the edge mask (w/o SR, i.e., $\lambda_{\mathrm{KL}}=0$) reduces AUROC and produces denser, less faithful explanations. Omitting frequency-domain features (w/o FFT) yields a marked drop, confirming the utility of spectral information for brain dynamics. Collectively, these results support the indispensability of each module in \textbf{\textit{SeeGraph}}.

% As an example of illustrating the model's explainability method, we visualized the key brain connectivity results on the SZPH dataset, showing how the model learns to capture and integrate temporal dependencies and functional connectivity across brain regions. Illustrated in in Figure~\ref{fig:AD_vs_HC}, AD patients exhibit a highly centralized connectivity pattern dominated by the temporal lobe. In contrast, HC subjects display a more distributed organization, highlighting frontal–temporal and parietal–occipital pathways. These findings suggest temporal-lobe-centered centralization in AD, versus balanced integration in HC.

\begin{figure}[h]
    \centering
    \includegraphics[width=0.49\textwidth]{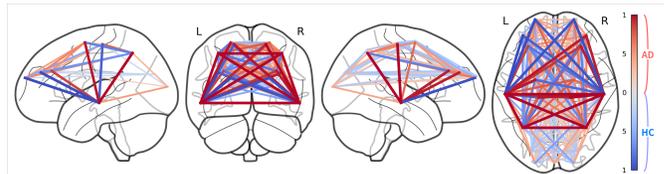} % 控制宽度
    % \caption{AD–HC functional connection differences.}
    \caption{\textbf{\textit{SeeGraph}} localizes salient fronto-temporal subnetworks via a node-guided sparse mask. AD exhibits temporal-centered ipsilateral clustering (red), while HC shows more interhemispheric and bilateral integration (blue); darker colors indicate higher salience.}
    \label{fig:AD_vs_HC}
\end{figure}

To illustrate the model’s explainability, we visualize disease-relevant brain connectivity on the SZPH cohort, showing how \textbf{\textit{SeeGraph}} captures and integrates temporal dependencies and interregional functional coupling.
As shown in Fig.~\ref{fig:AD_vs_HC}, disease-relevant edges concentrate in fronto-temporal circuitry. AD exhibits temporal-lobe–centered clustering, whereas HC retains more interhemispheric bridges and a more symmetric bilateral fronto-temporal integration.

\vspace{-0.5cm}
\section{Conclusion}
% In this work, we proposed  \textbf{\textit{SeeGraph}}, an interpretable spatio-temporal graph neural network for EEG-based Alzheimer’s disease diagnosis. By constructing dynamic brain networks and incorporating stochastic attention with an information bottleneck,  \textbf{\textit{SeeGraph}} yields compact connectivity patterns that enhance both robustness and interpretability. Experiments on two EEG datasets demonstrate state-of-the-art performance and clinically consistent frequency-band patterns, highlighting the potential of  \textbf{\textit{SeeGraph}} for transparent and reliable neurodegenerative disease analysis.

This paper presents \textbf{\textit{SeeGraph}}, a sparse-explanatory dynamic EEG-graph network addressing the dual challenges of robustness and interpretability in dementia diagnosis. By modeling time-varying brain networks, \textbf{\textit{SeeGraph}} jointly captures regional oscillations and interregional coupling, extracts clinically meaningful subnetworks through a node-guided sparse masking mechanism. The model integrates spectral and topological cues via a dual-trajectory encoder and Laplacian-based positional encoding, supporting transparent diagnostic reasoning. Experiments on a public and a clinical EEG cohort show that \textbf{\textit{SeeGraph}} achieves superior performance under clean and noisy conditions. Ablation and interpretability analyses validate each component and the alignment of extracted patterns with clinical findings. \textbf{\textit{SeeGraph}} thus offers a robust and explainable framework for EEG-based neurodegenerative disease analysis.
\vfill\pagebreak

% \section{ACKNOWLEDGEMENTS}
% This work was supported by the Shenzhen Science and Technology Program (Grant Nos. RCBS20231211090800003, ZDSYS2023062-
% -6091203008).
% \vspace{-0.5cm}

% \section{REFERENCES}
\label{sec:refs}

% List and number all bibliographical references at the end of the
% paper. The references can be numbered in alphabetic order or in
% order of appearance in the document. When referring to them in
% the text, type the corresponding reference number in square
% brackets as shown at the end of this sentence \cite{C2}. An
% additional final page (the fifth page, in most cases) is
% allowed, but must contain only references to the prior
% literature.

% Please follow the IEEE Citation Guidelines, \url{https://ieee-dataport.org/sites/default/files/analysis/27/IEEE\%20Citation\%20Guidelines.pdf} for formatting of references.

% \bibliographystyle{IEEEbib}
% \bibliography{strings,refs}

{\ninept
\bibliographystyle{IEEEbib}
\bibliography{strings,refs}
% }

\end{document}